\documentclass[superscriptaddress,english,floatfix,aps,pra,amsmath,amssymb,longbibliography,twocolumn]{revtex4-2}
\usepackage[T1]{fontenc}
\usepackage[utf8]{inputenc}
\usepackage{babel}
\usepackage{comment}
\usepackage{color}
\usepackage{times}
\usepackage{epsfig}
\usepackage{subfigure}
\usepackage{amsmath}
\usepackage{amssymb}
\usepackage{graphicx}
\usepackage[colorlinks=true]{hyperref}
\hypersetup{citecolor=blue,linkcolor=blue,urlcolor=blue}

\begin{document}

\title{Curvature effects on the regimes of the lateral van der Waals force}
\author{Alexandre P. Costa}
\email{alexandre.pereira.costa@icen.ufpa.br}
\affiliation{Faculdade de F\'{i}sica, Universidade Federal do Par\'{a}, 66075-110, Bel\'{e}m, Par\'{a}, Brazil}

\author{Lucas Queiroz}
\email{lucas.queiroz@ifpa.edu.br}
\affiliation{Instituto Federal de Educa\c{c}\~{a}o, Ci\^{e}ncia e Tecnologia do Par\'{a}, Campus Rural de Marab\'{a}, 68508-970, Marab\'{a}, Par\'{a}, Brazil}
\affiliation{Faculdade de F\'{i}sica, Universidade Federal do Par\'{a}, 66075-110, Bel\'{e}m, Par\'{a}, Brazil}

\author{Danilo T. Alves}
\email{danilo@ufpa.br}
\affiliation{Faculdade de F\'{i}sica, Universidade Federal do Par\'{a}, 66075-110, Bel\'{e}m, Par\'{a}, Brazil}

\date{\today}

\begin{abstract}
Recently, it has been shown that, under the action of the lateral van der Waals (vdW) force due to a perfectly conducting corrugated plane, a neutral anisotropic polarizable particle in vacuum can be attracted not only to the nearest corrugation peak but also to a valley or an intermediate point between a peak and a valley, with such behaviors called the peak, valley, and intermediate regimes, respectively.
%
In the present paper, we calculate the vdW interaction between a polarizable particle and a grounded conducting corrugated cylinder, and investigate how the effects of the curvature of the cylinder affect the occurrence of the mentioned regimes.
%
%
%
%
\end{abstract}

%

\maketitle

\section{Introduction}
\label{sec-intro}

A recent investigation of the van der Waals (vdW) interaction between a neutral polarizable particle and a corrugated plane, using analytical calculations beyond the proximity force approximation (PFA), led to the prediction that, under the action of the lateral van der Waals (vdW) force, an anisotropic particle can be attracted not only toward the nearest corrugation peak, but also to a valley, or to an intermediate point between a peak and valley, with such behaviors called as peak, valley and intermediate regimes, respectively  \cite{Nogueira-PRA-2021}.
%
%
%
%
In Ref. \cite{Queiroz-PRA-2021}, it was discussed how these regimes are affected by the consideration of nondispersive dielectric media, obtaining a first estimate about their behavior in the presence of dielectrics.
In Ref. \cite{Queiroz-PRA-2024}, it was shown how the occurrence of the peak, valley and intermediate regimes is affected by the consideration of realistic dielectric properties for the surface and also of the retardation in the interaction.

%
In the present paper, we discuss the influence of another factor on the occurrence of the regimes of the lateral force, namely, the influence of curvature on the corrugated surface.
In other words, we investigate how modifications on the geometric properties of the surface itself on which the corrugation occurs affect the mentioned regimes.
For this, we start considering the corrugations occurring on the surface of a grounded conducting cylinder and calculate the vdW interaction between a polarizable particle and such grounded conducting corrugated cylinder.
To compute this interaction, we start developing a generalization of the perturbative calculation discussed in Ref. \cite{Clinton-PRB-1985} to take into account the cylindrical geometry, and calculate the electrostatic potential of a charge in the presence of a grounded conducting corrugated cylinder.
In the sequence, we calculate the vdW interaction between a polarizable particle and a grounded conducting corrugated cylinder, and investigate the effects of the curvature of the cylinder on the regimes of the lateral vdW force discussed in Ref. \cite{Nogueira-PRA-2021}.

The paper is organized as follows. 
In Sec. \ref{sec-used-approach-regimes-curvatura}, we present the approach used to compute the vdW interaction between a neutral particle and a corrugated cylinder, and we discuss some applications.
In Sec. \ref{sec-vdw-interaction-regimes-curvatura}, we compute the vdW interaction for the mentioned situation, and we apply our results to the case of a sinusoidal corrugation.
In Sec. \ref{sec-final-regimes-curvatura}, we present our final comments.

\section{Point charge in the presence of a corrugated cylinder}
\label{sec-used-approach-regimes-curvatura}

\subsection{The solution of the Poisson's equation}

\begin{figure}
	\centering
	\subfigure[\label{fig:normal-cylinder-outside}]{\epsfig{file=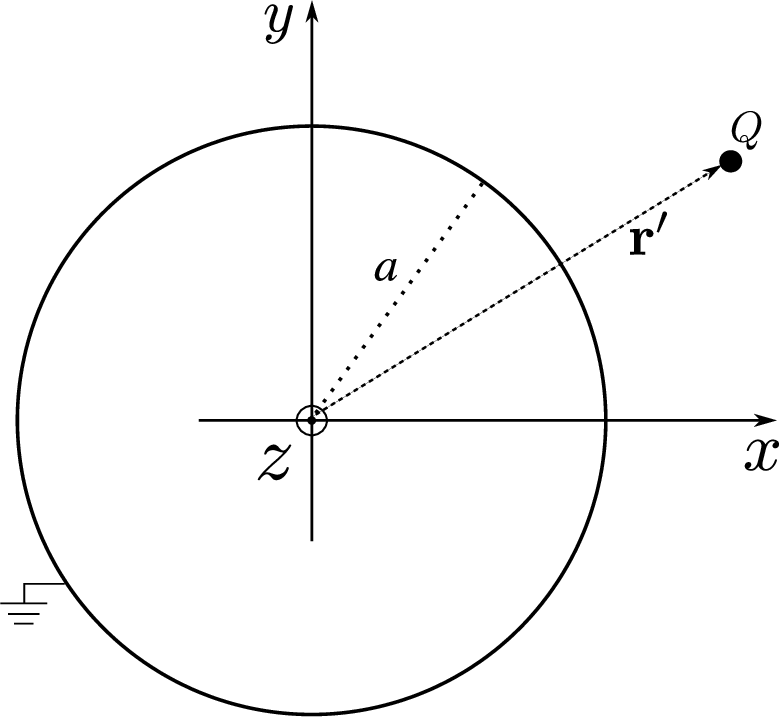, width=0.38 \linewidth}}
	\hspace{2mm}
	\subfigure[\label{fig:corrugated-cylinder-outside}]{\epsfig{file=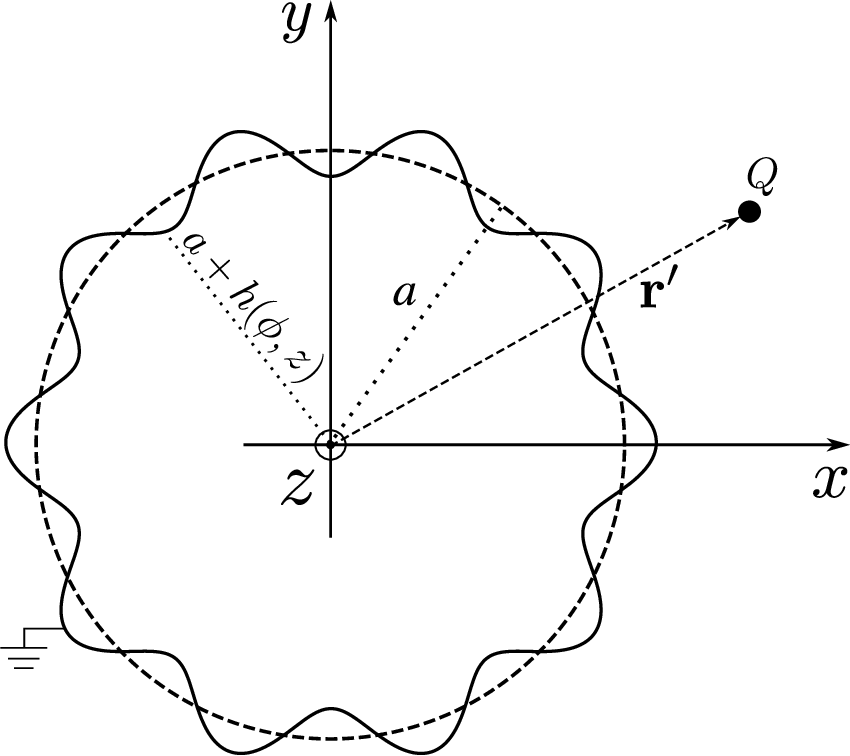, width=0.4 \linewidth}}
	\caption{ Ilustration of a charge $Q$, located at $\textbf{r}^{\prime}=\rho^{\prime}\boldsymbol{\hat{\rho}}+\phi^{\prime}\boldsymbol{\hat{\phi}}+z^{\prime}\mathbf{\hat{z}}$
		(with $\rho^\prime>a$), outside of two different cylinders.
		(a) A grounded conducting cylinder with radius  $a$. 
		(b) A grounded conducting corrugated cylinder, with the corrugation profile is described by $\rho=a+h(\phi,z)$. 
	}
\end{figure}

Let us start considering the problem of a point charge $Q$, located at the position $\textbf{r}^{\prime}=\rho^{\prime}\boldsymbol{\hat{\rho}}+\phi^{\prime}\boldsymbol{\hat{\phi}}+z^{\prime}\mathbf{\hat{z}}$, and interacting with an infinite grounded conducting cylinder with radius $a$ [see Fig. \ref{fig:normal-cylinder-outside}].
The Green's function related to the Poisson's equation for this problem is known from Ref. \cite{Hernandes-JE-2005}, and is given by
\begin{align}
G\left(\bf{r},\bf{r}^{\prime}\right)=2\sum_{j}\int \frac{dk}{2\pi} e^{ik\left(z-z^{\prime}\right)}e^{ij\left(\phi-\phi^{\prime}\right)} K_{j}\left(\left|k\right|\rho_{>}\right)\nonumber\\
\times\left[I_{j}\left(\left|k\right|\rho_{<}\right)-K_{j}\left(\left|k\right|\rho_{<}\right)\frac{I_{j}\left(\left|k\right|a\right)}{K_{j}\left(\left|k\right|a\right)}\right],
	\label{eq:solu-hernandes}
\end{align}
where $\rho_{<}$ and $\rho_{>}$ refer, respectively, to the smaller and the larger between $\rho$ and $\rho^{\prime}$, and, to avoid overloading, we are using $\sum_{j}\int \to \sum_{j=-\infty}^{+\infty}\int_{-\infty}^{+\infty}$.

Now, let us consider the previous problem and introduce in the cylinder surface a corrugation described by $\rho = a+h(\theta,z)$, where $h(\theta,z)$ describes a suitable modification 
on the cylinder surface, as shown in Fig. \ref{fig:corrugated-cylinder-outside}.
In this situation, the problem consists of finding the solution of Poisson's equation 
\begin{equation}
	\label{eq:Poisson}
	\nabla^{2}G\left(\bf{r},\bf{r}^{\prime}\right)=-\frac{4\pi}{\rho}\delta\left(\rho-\rho^{\prime}\right)\delta\left(\phi-\phi^{\prime}\right)\delta\left(z-z^{\prime}\right),
\end{equation}
under the boundary conditions
\begin{align}
	\left.G\left(\textbf{r},\textbf{r}^{\prime}\right)\right|_{\rho=a+ \varepsilon h\left(\phi,z\right)}&=0,
	\label{eq:bc-cyl-ext-epsilon}\\
	\left.G\left(\textbf{r},\textbf{r}^{\prime}\right)\right|_{|\textbf{r}|\gg |\textbf{r}^\prime|}&\to 0,
	\label{eq:bc-cyl-ext-infinity}
\end{align}
where $\varepsilon$ is an arbitrary auxiliary parameter such that $0\leq \varepsilon \leq 1$, where for $\varepsilon=0$ we recover a cylinder with radius $a$, and for $\varepsilon=1$ we recover the corrugated cylinder described by $h(\phi,z)$.
We look for a perturbative solution for the Green's function which is written in terms of the parameter $\varepsilon$ as 
\begin{equation}
\label{eq:green-gen}
G(\textbf{r},\textbf{r}^{\prime})=G^{(0)}(\textbf{r},\textbf{r}^{\prime})+\sum_{n=1}^{\infty} \varepsilon^{n} G^{(n)}(\textbf{r},\textbf{r}^{\prime}),
\end{equation}   
where $G^{(0)}(\textbf{r},\textbf{r}^{\prime})$ is the unperturbed solution [corresponding to the situation shown in Fig. \ref{fig:normal-cylinder-outside}], and $G^{(n)}(\textbf{r},\textbf{r}^{\prime})$ are the perturbative corrections to $G^{(0)}$ due to the surface corrugation.
Substituting Eq. \eqref{eq:green-gen} into Eq. \eqref{eq:bc-cyl-ext-epsilon}, and expanding in powers of $\varepsilon h(\phi,z)$, we obtain boundary conditions for $G^{(0)}(\textbf{r},\textbf{r}^{\prime})$ and $G^{(n)}(\textbf{r},\textbf{r}^{\prime})$ at $\rho=a$, which are given by
\begin{align}
G^{\left(0\right)}\left({\bf r},{\bf r^{\prime}}\right)\vert_{\rho=a}&=0, \label{eq:bc-G0}
\\
G^{\left(n\right)}\left({\bf r},{\bf r^{\prime}}\right)\vert_{\rho=a}&=-\sum_{m=1}^{n}\left[h(\phi,z)\right]^{m}\frac{1}{m!}\nonumber\\
&\times\frac{\partial^{m}}{\partial\rho^{m}}G^{\left(n-m\right)}\left({\bf r},{\bf r^{\prime}}\right)\vert_{\rho=a}. \label{eq:bc-Gn}
\end{align}

In order to solve Eq. \eqref{eq:Poisson}, it is convenient to write the Green's function as \cite{Jackson-Electrodynamics-1998}
\begin{equation}
G\left(\mathbf{r},\mathbf{r}^{\prime}\right)=\sum_{j}\int\frac{dk}{2\pi}\widetilde{G}_{j}\left(\rho,k,\mathbf{r}^{\prime}\right)e^{ikz}e^{ij\phi},
\label{eq:green-fourier}
\end{equation}
where
\begin{equation}
\widetilde{G}_{j}\left(\rho,k,\mathbf{r}^{\prime}\right)=\int_{0}^{2\pi}\frac{d\phi}{2\pi}\int dzG\left(\mathbf{r},\mathbf{r}^{\prime}\right)e^{-ikz}e^{-ij\phi}.
\label{eq:Fourier-inversa}
\end{equation}
Similarly, we can write the $\phi$ and $z$ delta functions as
\begin{align}
\delta\left(\phi-\phi^{\prime}\right)&	=\frac{1}{2\pi}\sum_{j}e^{ij\left(\phi-\phi^{\prime}\right)},  \\
\delta\left(z-z^{\prime}\right)&	=\int\frac{dk}{2\pi}e^{ik\left(z-z^{\prime}\right)}.
\end{align}
In this way, by using these equations into Eq. \eqref{eq:Poisson}, one obtains
%
\begin{align}
\left[\frac{1}{\rho}\frac{\partial}{\partial\rho}\left(\rho\frac{\partial}{\partial\rho}\right)-\left(\frac{j^{2}}{\rho^{2}}+k^{2}\right)\right]\widetilde{G}_{j}\left(\rho,k,{\bf r^{\prime}}\right) =-\frac{2}{\rho}\delta\left(\rho-\rho^{\prime}\right)\nonumber\\
\times e^{-i(j\phi^{\prime}+kz^{\prime})}.
\label{eq:poisson-transform}
\end{align}
We can also use Eq. \eqref{eq:green-fourier} into Eqs. \eqref{eq:green-gen}-\eqref{eq:bc-Gn}, which leads to
\begin{equation}
	\label{eq:green-transform}
	\widetilde{G}_{j}\left(\rho,k,{\bf r^{\prime}}\right)=\widetilde{G}_{j}^{(0)}\left(\rho,k,{\bf r^{\prime}}\right)+\sum_{n=1}^{\infty}\varepsilon^{n}\widetilde{G}_{j}^{(n)}\left(\rho,k,{\bf r^{\prime}}\right).
\end{equation}
and 
\begin{align}
\widetilde{G}_{j}^{\left(0\right)}\left(\rho,k,\bf{r}^{\prime}\right)\vert_{\rho=a}=&0, \label{eq:bc-G0-transform}
\\
\widetilde{G}_{j}^{\left(n\right)}\left(\rho,k,\bf{r}^{\prime}\right)\vert_{\rho=a}=&-\sum_{m=1}^{n}\sum_{j^{\prime}}\int\frac{dk^{\prime}}{2\pi} \widetilde{h}_{m,j,j^{\prime}}\left(k-k^{\prime}\right)	\frac{1}{m!}\nonumber\\
&\times\frac{\partial^{m}}{\partial\rho^{m}}\widetilde{G}_{j^{\prime}}^{\left(n-m\right)}\left(\rho,k^{\prime},\bf{r}^{\prime}\right)\vert_{\rho=a}, \label{eq:bc-Gn-transform}
\end{align}
where
\begin{align}
\label{eq:func-h-infinit}
\widetilde{h}_{m,j,j^{\prime}}\left(k-k^{\prime}\right)=\int_{0}^{2\pi}\frac{d\phi}{2\pi}\int dze^{-iz\left(k-k^{\prime}\right)}\nonumber\\
\times e^{-i\phi\left(j-j^{\prime}\right)}\left[h(\phi,z)\right]^{m}.
\end{align}

Substituting Eq. \eqref{eq:green-transform} into Eq. \eqref{eq:poisson-transform}, we obtain differential equations for $\widetilde{G}_{j}^{(0)}\left(\rho,k,{\bf r^{\prime}}\right)$ and $\widetilde{G}_{j}^{(n)}\left(\rho,k,{\bf r^{\prime}}\right)$, which are given by
\begin{gather}
\left[\frac{1}{\rho}\frac{\partial}{\partial\rho}\left(\rho\frac{\partial}{\partial\rho}\right)-\left(\frac{j^{2}}{\rho^{2}}+k^{2}\right)\right]\widetilde{G}_{j}^{\left(0\right)}\left(\rho,k,\bf{r}^{\prime}\right)\nonumber\\=-\frac{2}{\rho}\delta\left(\rho-\rho^{\prime}\right)
\times e^{-i(j\phi^{\prime}+kz^{\prime})}, \label{eq:G0-charge}
	\\
\left[\frac{1}{\rho}\frac{\partial}{\partial\rho}\left(\rho\frac{\partial}{\partial\rho}\right)-\left(\frac{j^{2}}{\rho^{2}}+k^{2}\right)\right]\widetilde{G}_{j}^{\left(n\right)}\left(\rho,k,\bf{r}^{\prime}\right)=0\;(n\geq1).\label{eq:Gn-charge}
\end{gather}
The solution of Eq. \eqref{eq:G0-charge}, with boundary conditions given by Eqs. \eqref{eq:bc-G0-transform} and
\begin{equation}
	\left.\widetilde{G}_{j}^{(0)}\left(\rho,k,{\bf r^{\prime}}\right)\right|_{\rho\gg\rho^{\prime}}\to0,
\end{equation}
can be written as
\begin{align}
\label{eq:solu-G0-exterior}
\widetilde{G}_{j}^{\left(0\right)}\left(\rho,k,\bf{r}^{\prime}\right)=&2e^{-ij\phi^{\prime}}e^{-ikz^{\prime}}K_{j}\left(\left|k\right|\rho_{>}\right)\nonumber\\ &\times\left[I_{j}\left(\left|k\right|\rho_{<}\right)-K_{j}\left(\left|k\right|\rho_{<}\right)\frac{I_{j}\left(\left|k\right|a\right)}{K_{j}\left(\left|k\right|a\right)}\right],
\end{align} 
where $I_{j}$ and $K_{j}$ are the modified Bessel functions of first and second kind, respectively.
In this way, the solution for $G^{\left(0\right)} \left(\bf{r},\bf{r}^{\prime}\right)$ can be obtained from Eq. \eqref{eq:green-fourier}, so that one finds that
\begin{align}
G^{\left(0\right)}\left(\bf{r},\bf{r}^{\prime}\right)=2\sum_{j}\int \frac{dk}{2\pi} e^{ik\left(z-z^{\prime}\right)}e^{ij\left(\phi-\phi^{\prime}\right)} K_{j}\left(\left|k\right|\rho_{>}\right)\nonumber\\
\times\left[I_{j}\left(\left|k\right|\rho_{<}\right)-K_{j}\left(\left|k\right|\rho_{<}\right)\frac{I_{j}\left(\left|k\right|a\right)}{K_{j}\left(\left|k\right|a\right)}\right],
	\label{eq:solu-G0-coordinates}
\end{align}
which, as expected, coincides with Eq. \eqref{eq:solu-hernandes}.

The solution of Eq. \eqref{eq:Gn-charge} with the boundary conditions given by Eq. \eqref{eq:bc-Gn-transform} and
\begin{equation}
\left. \widetilde{G}_{j}^{\left(n\right)}\left(\rho,k,\bf{r}^{\prime}\right) \right| _{\rho\gg\rho^{\prime}}\to0,
\end{equation}
can be written as
\begin{align}
\label{eq:solu-Gn-exterior}
\widetilde{G}_{j}^{\left(n\right)}\left(\rho,k,\bf{r}^{\prime}\right)=&-\sum_{m=1}^{n}\sum_{j^{\prime}}\frac{K_{j}\left(\left|k\right|\rho\right)}{K_{j}\left(\left|k\right|a\right)} \int\frac{dk^{\prime}}{2\pi}\widetilde{h}_{m,j,j^{\prime}}\left(k-k^{\prime}\right)\nonumber\\
&\times\frac{1}{m!} \left[\frac{\partial^{m}}{\partial\rho^{m}}\widetilde{G}_{j^{\prime}}^{\left(n-m\right)}\left(\rho,k^{\prime},\bf{r}^{\prime}\right)\right]_{\rho=a}.
\end{align}
From Eq. \eqref{eq:green-fourier}, one can find the solution for $G^{\left(n\right)} \left(\bf{r},\bf{r}^{\prime}\right)$, which can be written as
\begin{align}
\label{eq:gn-cilindro}
G^{(n)}(\mathbf{r},\mathbf{r}^{\prime})&=-\sum_{m=1}^{n}\sum_{j,j^{\prime}}\int\frac{dk}{2\pi}\int\frac{dk^{\prime}}{2\pi}\frac{K_{j}(|k|\rho)}{K_{j}(|k|a)}e^{ikz}e^{ij\phi}\nonumber\\
&\times\frac{\widetilde{h}_{m,j,j^{\prime}}(k-k^{\prime})}{m!}\left[\frac{\partial^{m}}{\partial\rho^{m}}\widetilde{G}_{j^{\prime}}^{(n-m)}(\rho,k^{\prime},\mathbf{r}^{\prime})\right]_{\rho=a}.
\end{align}
Thus, the solution of Eq. \eqref{eq:Poisson} for the problem of a point charge in the presence of a corrugated cylinder is given by Eq. \eqref{eq:green-gen} (with $\varepsilon=1$), with $ G^{(0)} $ and $ G^{(n)} $ given by Eqs. \eqref{eq:solu-G0-coordinates} and \eqref{eq:gn-cilindro}, respectively.
Next, we apply the obtained results to compute the interaction between the charge and the corrugated cylinder.
%


\subsection{Interaction energy for a general corrugated cylinder}

The interaction between a unitary point charge and the corrugated cylinder shown in Fig. \ref{fig:corrugated-cylinder-outside} is given, in Gaussian units, by
\begin{equation}
U\left(\mathbf{r}\right)=\left[G\left(\mathbf{r},\mathbf{r^{\prime}}\right)-\frac{1}{\vert\mathbf{r}-\mathbf{r^{\prime}}\vert}\right]_{\mathbf{r}=\mathbf{r^{\prime}}},
\end{equation}
where $ G\left(\mathbf{r},\mathbf{r^{\prime}}\right) $ is given by Eq. \eqref{eq:green-gen} (with $\varepsilon=1$), with $ G^{(0)} $ and $ G^{(n)} $ given by Eqs. \eqref{eq:solu-G0-coordinates} and \eqref{eq:gn-cilindro}, respectively.
By using the following expansion for $1/\vert\mathbf{r}-\mathbf{r^{\prime}}\vert$ \cite{Jackson-Electrodynamics-1998}
\begin{align}
\frac{1}{\vert\mathbf{r}-\mathbf{r^{\prime}}\vert} &=\sum_{j}\int \frac{dk}{\pi} e^{ik\left(z-z^{\prime}\right)}e^{ij\left(\phi-\phi^{\prime}\right)} K_{j}\left(\left|k\right|\rho_{>}\right)I_{j}\left(\left|k\right|\rho_{<}\right),
\end{align}
we can write $U\left(\mathbf{r}\right)$ as 
\begin{equation}
U\left(\mathbf{r}\right)=\left[G_{H}^{\left(0\right)}\left({\bf r},{\bf r^{\prime}}\right)+\sum_{n=1}^{\infty}G^{\left(n\right)}\left({\bf r},{\bf r^{\prime}}\right)\right]_{\mathbf{r}=\mathbf{r^{\prime}}},
\end{equation}
where $ G_{H}^{\left(0\right)}\left({\bf r,{\bf r^{\prime}}}\right) $ is given by
\begin{align}
G_{H}^{\left(0\right)}\left({\bf r,{\bf r^{\prime}}}\right)=&-\sum_{j}\int\frac{dk}{\pi}e^{ik\left(z-z^{\prime}\right)}e^{ij\left(\phi-\phi^{\prime}\right)}K_{j}\left(\left|k\right|\rho_{>}\right)\nonumber\\
&\times K_{j}\left(\left|k\right|\rho_{<}\right)\frac{I_{j}\left(\left|k\right|a\right)}{K_{j}\left(\left|k\right|a\right)}.
\end{align}
By performing $ \mathbf{r}=\mathbf{r^{\prime}} $, we obtain
\begin{equation}
\label{sol-energia-geral-carga}
U\left(\mathbf{r}\right)=U^{\left(0\right)}({\bf r})+\sum_{n=1}^{\infty}U^{(n)}(\mathbf{r}),
\end{equation}
where
\begin{equation}
U^{\left(0\right)}({\bf r})=-\sum_{j}\int\frac{dk}{\pi}\frac{I_{j}\left(\left|k\right|a\right)}{K_{j}\left(\left|k\right|a\right)}\left[K_{j}\left(\left|k\right|\rho\right)\right]^{2},
\end{equation}
is the interaction between the charge and an infinite grounded conducting cylinder [as shown in Fig. \ref{fig:normal-cylinder-outside}]. 
The functions
\begin{align}
\label{eq:sol-geral-un-carga}
U^{(n)}(\mathbf{r})=&-\sum_{m=1}^{n}\sum_{j,j^{\prime}}\int\frac{dk}{2\pi}\int\frac{dk^{\prime}}{2\pi}\frac{K_{j}(|k|\rho)}{K_{j}(|k|a)}\frac{\widetilde{h}_{m,j,j^{\prime}}(k-k^{\prime})}{m!}\nonumber\\
&\times\left[\frac{\partial^{m}}{\partial\rho^{m}}\widetilde{G}_{j^{\prime}}^{(n-m)}(\rho,k^{\prime},\mathbf{r})\right]_{\rho=a}e^{ikz}e^{ij\phi},
\end{align}
with $ \widetilde{h}_{m,j,j^{\prime}} $ and $ \widetilde{G}_{j^{\prime}}^{(n-m)} $ given by Eqs. \eqref{eq:func-h-infinit} and \eqref{eq:solu-Gn-exterior}, respectively, are the corrections of $ U^{\left(0\right)} $ due to the presence of corrugations in the cylinder surface [as shown in Fig. \ref{fig:corrugated-cylinder-outside}].
Note that, since $ U\left(\mathbf{r}\right) $ depends on $\phi$ or $z$, a lateral force (parallel to the surface of the reference cylinder shown in Fig. \ref{fig:normal-cylinder-outside}) acting on the charge arises, and it occurs due to the presence of corrugations on the cylinder surface.
Moreover, the dependence of $ U\left(\mathbf{r}\right) $ on $\phi$ or $z$ belongs specifically to the perturbative corrections $U^{(n)}$.
Thus, since we are interested only in the behavior of the mentioned lateral force, hereafter we focus our attention only on $U^{(n)}$.

\subsection{Interaction energy for a cylinder with a sinusoidal corrugation along the $z$-direction}
\label{cap-2-sec-3-subsec-1}

\begin{figure}
\centering
\epsfig{file=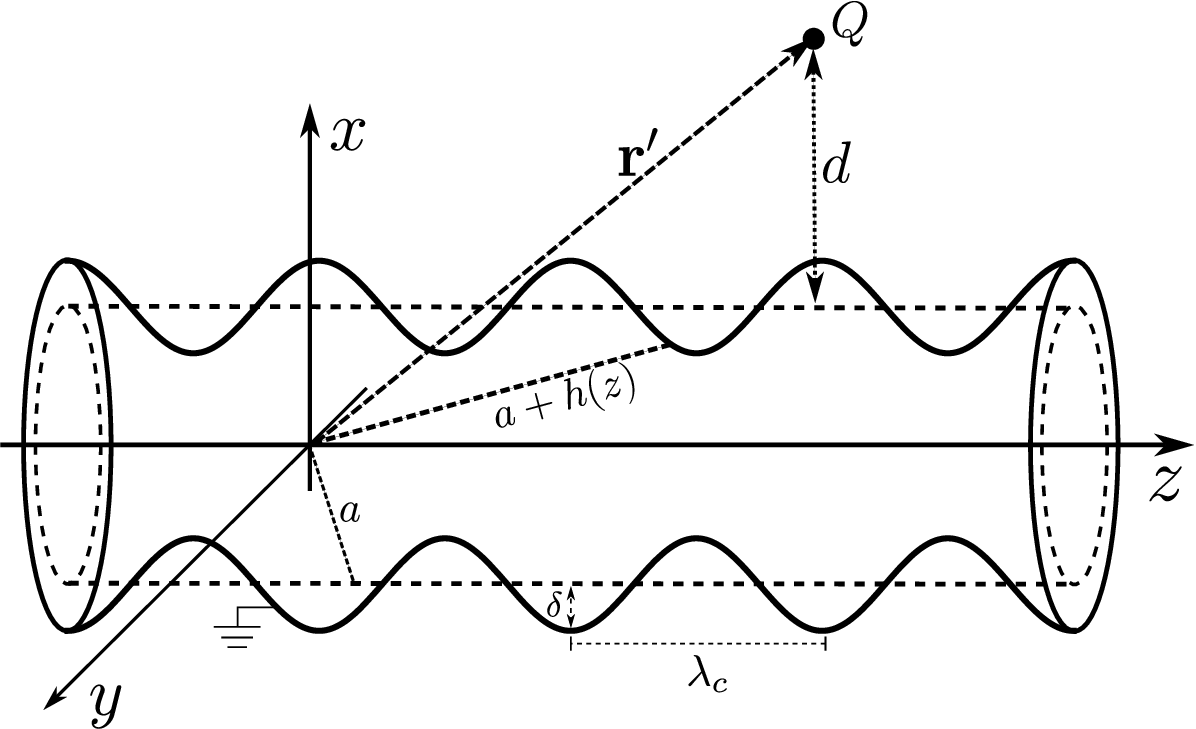, width=0.6 \linewidth}
\caption{
Illustration of a charge $Q$, located at $\mathbf{r}^{\prime}=\rho^{\prime}\boldsymbol{\hat{\rho}}+\phi^{\prime}\boldsymbol{\hat{\phi}}+z^{\prime}\mathbf{\hat{z}}$ [with $\rho^\prime>a+h\left(z\right)$], outside of a grounded conducting corrugated cylinder (solid line), whose corrugation profile is described by $h\left(z\right)=\delta\cos\left(k_c z\right)$.
The dashed line is the illustration of the cylinder without corrugation and with radius $a$.
}
\label{fig:corrugated-cylinder-z}
\end{figure}

\begin{figure}
\centering
\epsfig{file=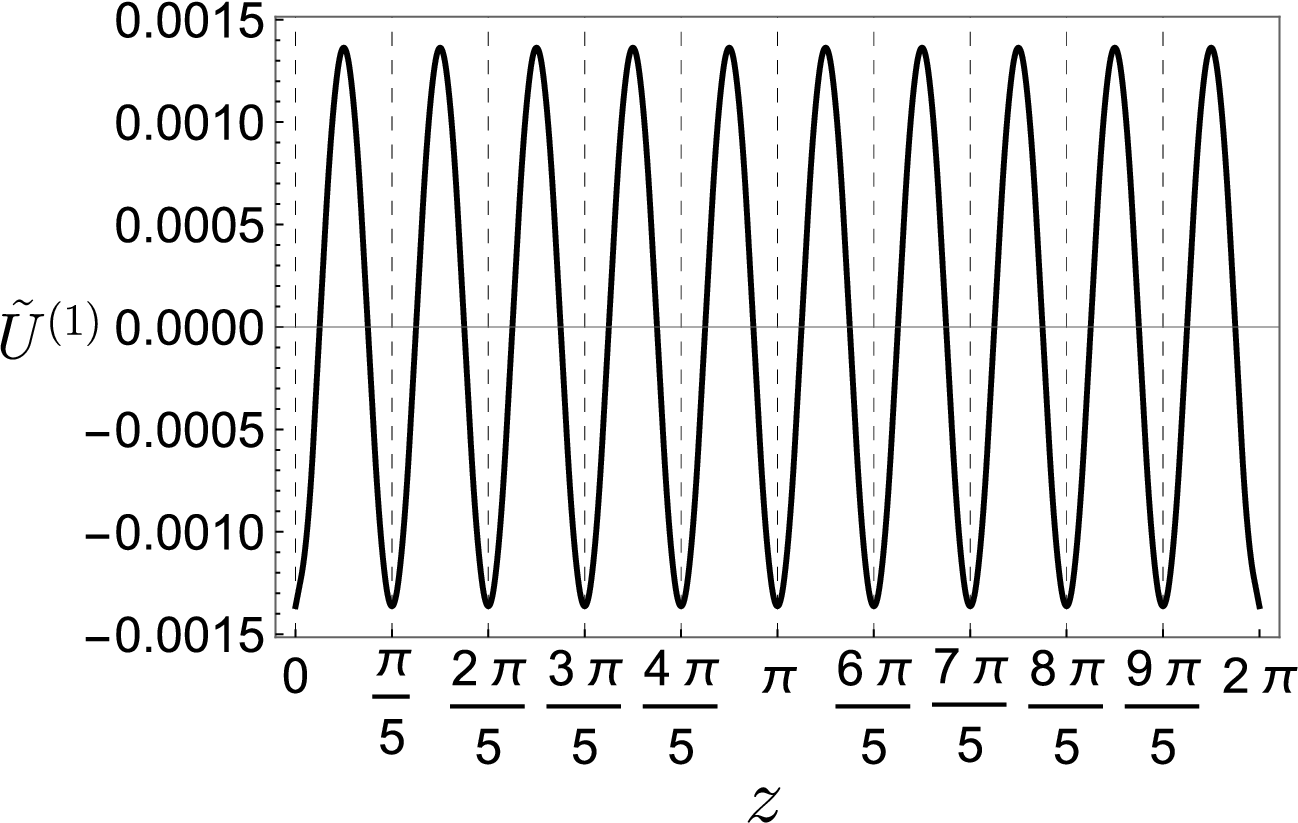,  width=0.56 \linewidth}
\caption{
Behavior of $\tilde{U}_{\text{cc-z}}^{\left(1\right)} = U_{\text{cc-z}}^{\left(1\right)}/(\delta/a\pi)$ versus $z$, with $a=1$, $d=1$ and $k_{c}=10$.
Each tick in the horizontal axis represents a corrugation peak.
}
\label{fig:sinusoidal-energy-z}
\end{figure}

Let us consider a sinusoidal corrugation profile described by 
\begin{equation}
	\label{eq:def-corrug-senoid-z}
	h\left(z\right)=\delta\cos\left(k_c z\right),
\end{equation}
where, $\delta$ is the amplitude of the corrugation and $ k_c = {2\pi}/{\lambda_{c}} $, with $\lambda_{c}$ being the wavelength of the corrugation (see Fig. \ref{fig:corrugated-cylinder-z}).
Let us also consider an approximate solution for this case which takes into account only the correction of $ n=1 $ in Eq. \eqref{sol-energia-geral-carga}.
In this way, from Eq. \eqref{eq:sol-geral-un-carga}, one obtains that $ U^{(1)} $ can be written, for the corresponding case, as
\begin{align}
U_{\text{cc-z}}^{\left(1\right)}\left(\mathbf{r}\right)=&-\frac{\delta}{a\pi}\cos\left(k_c z\right)\sum_{j}\int dk\frac{K_{j}\left[\left|k+k_c\right|\left(d+a\right)\right]}{K_{j}\left(\left|k+k_c\right|a\right)}\nonumber\\
&\times\frac{K_{j}\left[\left|k\right|\left(d+a\right)\right]}{K_{j}\left(\left|k\right|a\right)},
\label{eq:u1-carga-z}
\end{align}
where $d=\rho - a$ is the distance from the charge to the surface of the cylinder without the corrugations (see Fig. \ref{fig:corrugated-cylinder-z}), and the subscript ``cc-z'' refers to the result for a corrugated cylinder with corrugation occurring along the $z$-direction.
In Fig. \ref{fig:sinusoidal-energy-z}, we show the behavior of $\tilde{U}_{\text{cc-z}}^{\left(1\right)} = U_{\text{cc-z}}^{\left(1\right)}/(\delta/a\pi)$ versus $z$, where one can see that the minimum values of $U_{\text{cc-z}}^{\left(1\right)}$ are located over the corrugation peaks.
This means that the lateral force acting on the charge always attracts it to the nearest peak of the corrugation.
From Eq. \eqref{eq:u1-carga-z}, one sees that such behavior occurs for any value of $a$, $d$ and $k_c$, since the integrate and summation are positive for any value of these parameters, and all the dependence of $U_{\text{cc-z}}^{\left(1\right)}$ on $z$ is inserted in the cosine function.
%
%
%
%

\begin{figure}
\centering
\epsfig{file=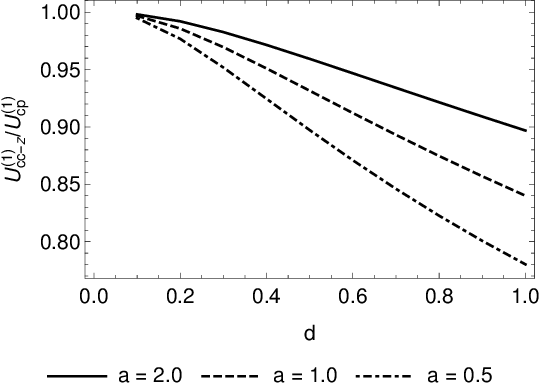,  width=0.56 \linewidth}
\caption{
Behavior of $ U_{\text{cc-z}}^{\left(1\right)}/U_{\text{cp}}^{\left(1\right)} $ versus $d$, for $k_{c}=10$ and $a = 0.5$ (dot-dashed line), $a=1$ (dashed line), and $a=2$ (solid line).
}
\label{fig:ratio-corrugation-z}
\end{figure}

It is also interesting to compare Eq. \eqref{eq:u1-carga-z} with the corresponding result for a corrugated plane, since it shows us the effect of the curvature of the surface on the lateral force.
From Ref. \cite{Clinton-PRB-1985}, one can find, perturbatively, the interaction between a point charge and a corrugated plane described by Eq. \eqref{eq:def-corrug-senoid-z}, whose first perturbative correction of the interaction is given by
\begin{equation}
\label{eq:U1_surface-z}
U_{\text{cp}}^{\left(1\right)}\left(\mathbf{r}\right)=-\frac{\delta}{4}k_c^{2}\cos\left(k_c z\right)K_{2}\left(k_c d\right),
\end{equation}
where the subscript ``cp'' refers to the result for a corrugated plane.
In Fig. \ref{fig:ratio-corrugation-z}, we show the behavior of the ratio $ U_{\text{cc-z}}^{\left(1\right)}/U_{\text{cp}}^{\left(1\right)} $ versus $d$, for different values of the cylinder radius $a$.
For $d \ll a $, one sees that $U_{\text{cc-z}}^{\left(1\right)} / U_{\text{cp}}^{\left(1\right)} \to 1$, which is expected, since as the charge approaches to the cylinder the curvature effects vanish.
On the other hand, for $d \sim a$, the presence of curvature on the corrugated surface becomes relevant and produces a weakening in $U^{\left(1\right)}$, which means that the lateral force for a corrugated cylinder is weaker than that for a corrugated plane.
%
%
%
%

\subsection{Interaction energy for a cylinder with a sinusoidal corrugation along the $\phi$-direction}

\begin{figure}
\centering
\epsfig{file=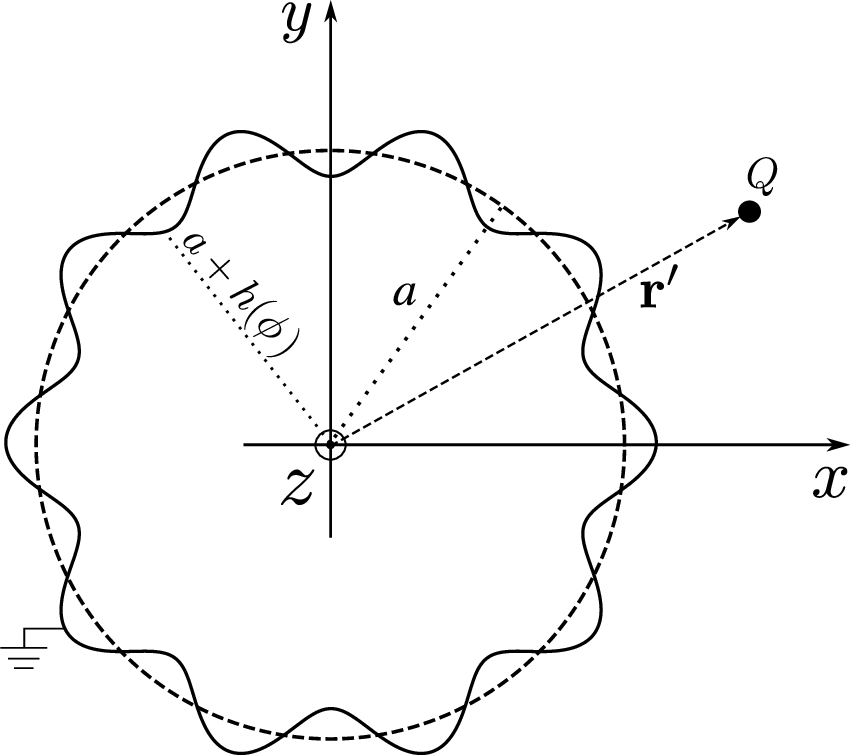, width=0.45 \linewidth}
\caption{
Illustration of a charge $Q$, located at $\mathbf{r}^{\prime}=\rho^{\prime}\boldsymbol{\hat{\rho}}+\phi^{\prime}\boldsymbol{\hat{\phi}}+z^{\prime}\mathbf{\hat{z}}$ [with $\rho^\prime>a+h\left(\phi\right)$], outside of a grounded conducting corrugated cylinder (solid line), whose corrugation profile is described by $h\left(\phi\right)=\delta\cos\left(k_c a \phi\right)$.
The dashed line is the illustration of the cylinder without corrugation and with radius $a$.
}
\label{fig:corrugated-cylinder-phi}
\end{figure}

\begin{figure}
\centering
\epsfig{file=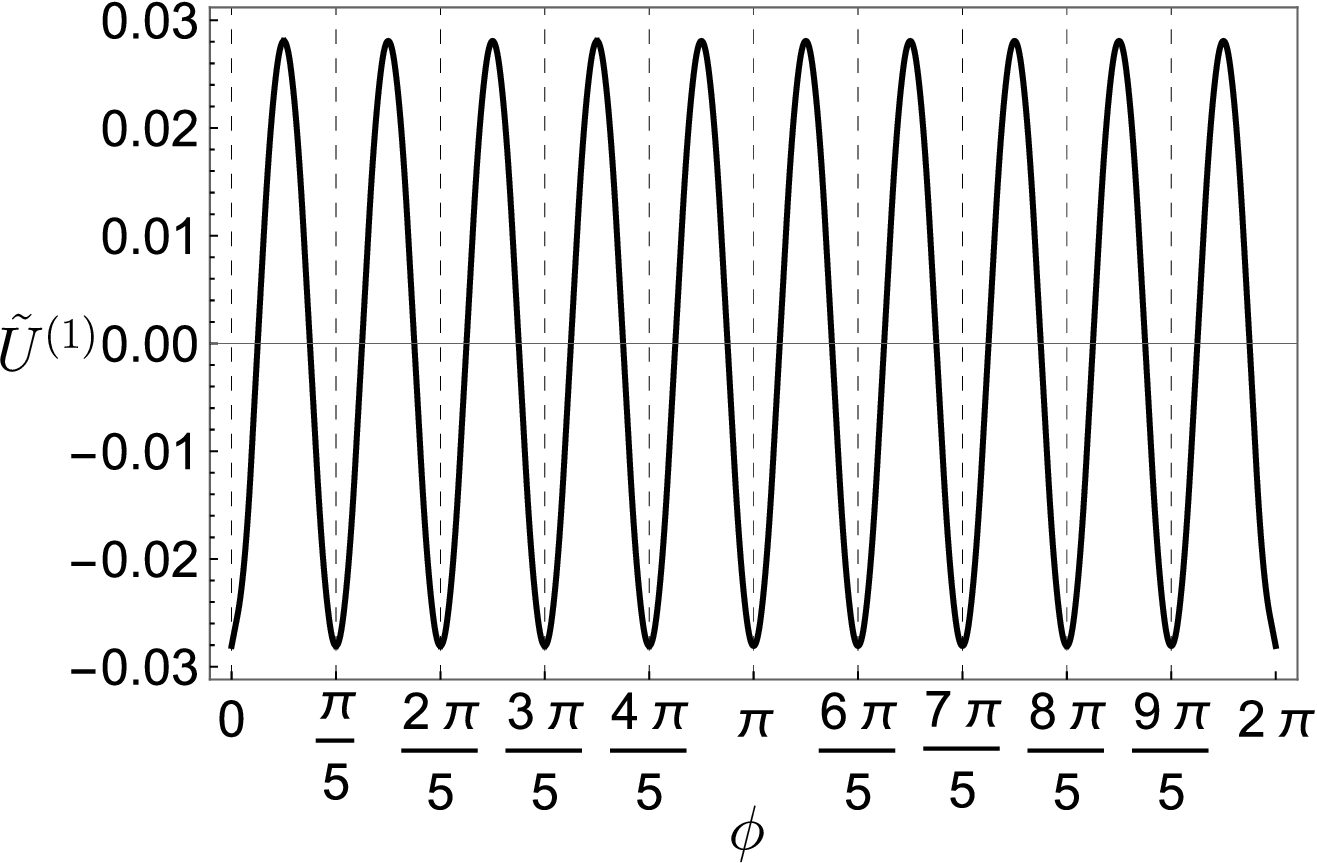,  width=0.56 \linewidth}
\caption{
Behavior of $\tilde{U}_{\text{cc-\ensuremath{\phi}}}^{\left(1\right)} = U^{\left(1\right)}_{\text{cc-}\phi} / \left(\delta /a\pi\right)$ versus $ \phi $, with $a=1$, $d=1$ and $k_{c}=10$.
Each tick on the horizontal axis represents a corrugation peak.
}
\label{fig:sinusoidal-energy-phi}
\end{figure}

Let us now consider a sinusoidal corrugation profile described by 
\begin{equation}
\label{eq:def-corrug-senoid}
h\left(\phi\right)=\delta\cos\left(k_c a\phi\right),
\end{equation}
where $\delta$ is the amplitude of the corrugation, $k_c=2\pi/\lambda_{c}$, with $\lambda_{c}$ being the wavelength of the corrugation and $a$ is the radius of the cylinder (see Fig. \ref{fig:corrugated-cylinder-phi}).
We remark that in this situation the number of oscillations of the corrugation is constrained to the length of the cylinder circumference, so that $\lambda_{c}$ must satisfy the constraint
\begin{equation}
\label{eq:nu}
N{\lambda_{c}}=2\pi a \Longrightarrow N = k_c a,
\end{equation}
where $N\in\mathbb{N}^*$. 
Let us also consider an approximate solution for this case which takes into account only the correction of $ n=1 $ in Eq. \eqref{sol-energia-geral-carga}.
In this way, from Eq. \eqref{eq:sol-geral-un-carga}, one obtains that $ U^{(1)} $ can be written, for the corresponding case, as
\begin{align}
\label{eq:u1-carga-phi}
U_{\text{cc-\ensuremath{\phi}}}^{\left(1\right)}\left(\mathbf{r}\right)=&-\frac{\delta}{a\pi}\cos\left(k_c a\phi\right)\sum_{j}\int dk\frac{K_{j}\left[\left|k\right|\left(d+a\right)\right]}{K_{j}\left(\left|k\right|a\right)}\nonumber\\
&\times\frac{K_{j+k_c a}\left[\left|k\right|\left(d+a\right)\right]}{K_{j+k_c a}\left(\left|k\right|a\right)}.
\end{align}
where the subscript ``cc-$\phi$'' refers to the result for a corrugated cylinder with corrugation occurring along the $\phi$-direction.
In Fig. \ref{fig:sinusoidal-energy-phi}, we show the behavior of $\tilde{U}_{\text{cc-\ensuremath{\phi}}}^{\left(1\right)} = U_{\text{cc-\ensuremath{\phi}}}^{\left(1\right)}/(\delta/a\pi)$ versus $\phi$, where one can see that the minimum values of $U_{\text{cc-\ensuremath{\phi}}}^{\left(1\right)}$ are located over the corrugation peaks.
This means that the lateral force acting on the charge always attracts it to the nearest peak of the corrugation.
%
%
Besides this, from Eq. \eqref{eq:u1-carga-phi}, one sees that such behavior occurs for any value of $a$, $d$ and $k_c$, since the integrate and summation are positive for any value of these parameters, and all the dependence of $U_{\text{cc-}\phi}^{\left(1\right)}$ on $\phi$ is inserted in the cosine function.

\begin{figure}
\centering
\epsfig{file=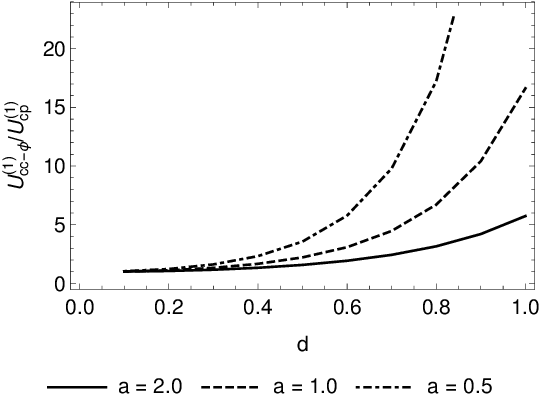,  width=0.56 \linewidth}
\caption{
Behavior of $U_{\text{cc-\ensuremath{\phi}}}^{\left(1\right)} / U_{\text{cp}}^{\left(1\right)}$ versus $d$ for a charge over a corrugation peak $(z=\phi=0)$. 
We are considering $k_{c}=10$ and $a = 0.5$ (dot-dashed line), $a=1$ (dashed line), and $a=2$ (solid line).
}
\label{fig:ratio-corrugation-phi}
\end{figure}

It is also interesting to compare Eq. \eqref{eq:u1-carga-phi} with Eq. \eqref{eq:U1_surface-z}, since it shows us the effect of the curvature of the surface on the lateral force.
In Fig. \ref{fig:ratio-corrugation-phi}, we show the behavior of the ratio $ U_{\text{cc-\ensuremath{\phi}}}^{\left(1\right)}/U_{\text{cp}}^{\left(1\right)} $ versus $d$, for different values of the cylinder radius $a$.
From this figure, for $d \ll a $, one sees the curvature effects vanishing, since $U_{\text{cc-\ensuremath{\phi}}}^{\left(1\right)}/U_{\text{cp}}^{\left(1\right)} \to 1$, and, for $d \sim a$, the presence of curvature on the corrugated surface becomes relevant, producing an amplification in the lateral force, different from the case in which the corrugation occurs in the $ z $-direction.

\section{The van der Waals interaction between a polarizable particle and a corrugated cylinder}
\label{sec-vdw-interaction-regimes-curvatura}

\subsection{Interaction energy for a general corrugated cylinder}

Let us consider a neutral polarizable particle, put at $\textbf{r}_0=\rho_0\boldsymbol{\hat{\rho}}+\phi_0\boldsymbol{\hat{\phi}}+z_0\mathbf{\hat{z}}$, in the presence of a grounded conducting corrugated cylinder with radius $a$, as illustrated in Fig. \ref{fig:corrugated-cylinder-outside}.
The vdW interaction $U_\text{vdW}$ between this particle and the corrugated cylinder can be computed by combining the Eberlein-Zietal formula from Ref. \cite{Eberlein-PRA-2007} with the solution for the Green’s function for the problem of a charge in the presence of a grounded conducting corrugated cylinder [Eqs. \eqref{eq:green-gen}, with $ G^{(0)} $ and $ G^{(n)} $ given by Eqs. \eqref{eq:solu-G0-coordinates} and \eqref{eq:gn-cilindro}].
The mentioned Eberlein-Zietal formula is given by
\begin{equation}
U_{\text{vdW}}(\mathbf{r}_0) = \frac{1}{8\pi\epsilon_{0}}\displaystyle\sum_{i,j}\langle \hat{d}_i \hat{d}_j\rangle\boldsymbol{ \nabla}_i\boldsymbol{ \nabla}_j' \left.G_H(\mathbf{r},\mathbf{r'})\right|_{\mathbf{r}=\mathbf{r'}=\mathbf{r}_0},
\label{eq:Eberlein_Zietalquantum}
\end{equation}
where $\hat{d}_i$ are the components of the dipole moment operator and $\langle \hat{d}_i \hat{d}_j\rangle$ is the expectation value of $\hat{d}_i \hat{d}_j$.
For simplicity, let us focus on an approximate solution for this case which takes into account only the correction of $ n=1 $ in Eq. \eqref{eq:gn-cilindro}.
Thus, by computing $G_H$ using 
\begin{equation}
G_H\left(\textbf{r},\textbf{r}^{\prime}\right) = G\left(\textbf{r},\textbf{r}^{\prime}\right) - \frac{1}{|\textbf{r}-\textbf{r}^{\prime}|}.
\label{eq:GH-exato}
\end{equation}
and substituting it into Eq. \eqref{eq:Eberlein_Zietalquantum}, one obtains that the vdW interaction can be written as $U_\text{vdW}\approx U^{(0)}_\text{vdW} + U^{(1)}_\text{vdW}$.
The first term $U^{(0)}_\text{vdW}$, which can be written as
\begin{equation}
\label{eq:energy-u0}
U_{\text{vdW}}^{(0)}\left(\mathbf{r}_{0}\right)=\frac{1}{8\pi\varepsilon_{0}}\left[\Xi_{\rho}\left\langle \widehat{d}_{\rho}^{2}\right\rangle +\Xi_{\phi}\left\langle \widehat{d}_{\phi}^{2}\right\rangle +\Xi_{z}\left\langle \widehat{d}_{z}^{2}\right\rangle \right],
\end{equation}
where
\begin{align}
\Xi_{\rho}&=-\frac{1}{\pi}\sum_{j}\int dk\frac{I_{j}\left(\left|k\right|a\right)}{K_{j}\left(\left|k\right|a\right)}\left[\partial_{\rho_{0}}K_{j}\left(\left|k\right|\rho_{0}\right)\right]^{2},
\\
\Xi_{\phi}&=-\frac{1}{\pi\rho^{2}_{0}}\sum_{j}\int  dk j^{2} \frac{I_{j}\left(\left|k\right|a\right)}{K_{j}\left(\left|k\right|a\right)}\left[K_{j}\left(\left|k\right|\rho_{0}\right)\right]^{2},
\\
\Xi_{z}&=-\frac{1}{\pi}\sum_{j}\int dk k^{2} \frac{I_{j}\left(\left|k\right|a\right)}{K_{j}\left(\left|k\right|a\right)}\left[K_{j}\left(\left|k\right|\rho_{0}\right)\right]^{2},
\end{align}
is the vdW interaction energy between the particle and the cylinder without corrugations, which was first obtained in Ref. \cite{Eberlein-PRA-2007}.
%
%
The second term, $ U^{(1)}_{\text{vdW}} $, is the first-order correction to $U^{(0)}_{\text{vdW}}$ due to the corrugation in the cylinder surface. 
Besides this, it is obtained by substituting $G^{(1)}$, given by Eq. \eqref{eq:gn-cilindro} with $ n=1 $, in the Eberlein-Zietal formula [Eq. \eqref{eq:Eberlein_Zietalquantum}], obtaining
\begin{align}
\label{eq:u-corrug-general}
U_{\text{vdW}}^{\left(1\right)}(\mathbf{r}_{0})=&\frac{1}{8\pi\epsilon_{0}}[\langle\hat{d}_{\rho}^{2}\rangle\mathcal{I}_{\rho\rho}+\langle\hat{d}_{\phi}^{2}\rangle\mathcal{I}_{\phi\phi}+\langle\hat{d}_{z}^{2}\rangle\mathcal{I}_{zz}\nonumber\\
&+\langle\hat{d}_{\rho}\hat{d}_{\phi}\rangle\mathcal{I}_{\rho\phi}+\langle\hat{d}_{\rho}\hat{d}_{z}\rangle\mathcal{I}_{\rho z}+\langle\hat{d}_{\phi}\hat{d}_{z}\rangle\mathcal{I}_{\phi z}],
\end{align}
where
%
\begin{align}
\mathcal{I}_{\rho\rho}=& -\frac{2}{a}\sum_{j,j^{\prime}}\int\frac{dk}{2\pi}\int\frac{dk^{\prime}}{2\pi}\widetilde{h}_{1,j,j^{\prime}}(k,k^{\prime})e^{iz(k-k^{\prime})}e^{i\phi(j-j^{\prime})}\nonumber\\
&\times\frac{\partial_{\rho}K_{j}(|k|\rho)}{K_{j}(|k|a)}\frac{\partial_{\rho}K_{j^{\prime}}(|k^{\prime}|\rho)}{K_{j^{\prime}}(|k^{\prime}|a)},\\
\mathcal{I}_{\phi\phi}=& -\frac{2}{a\rho^{2}}\sum_{j,j^{\prime}}\int\frac{dk}{2\pi}\int\frac{dk^{\prime}}{2\pi}\widetilde{h}_{1,j,j^{\prime}}(k,k^{\prime})jj^{\prime}e^{iz(k-k^{\prime})}\nonumber\\
&\times e^{i\phi(j-j^{\prime})}\frac{K_{j}(|k|\rho)}{K_{j}(|k|a)}\frac{K_{j^{\prime}}(|k^{\prime}|\rho)}{K_{j^{\prime}}(|k^{\prime}|a)},\\
\mathcal{I}_{zz}=& -\frac{2}{a}\sum_{j,j^{\prime}}\int\frac{dk}{2\pi}\int\frac{dk^{\prime}}{2\pi}\widetilde{h}_{1,j,j^{\prime}}(k,k^{\prime})kk^{\prime}e^{iz(k-k^{\prime})}\nonumber\\
&\times e^{i\phi(j-j^{\prime})}\frac{K_{j}(|k|\rho)}{K_{j}(|k|a)}\frac{K_{j^{\prime}}(|k^{\prime}|\rho)}{K_{j^{\prime}}(|k^{\prime}|a)},\\
\mathcal{I}_{\rho\phi}=& -\frac{2i}{a\rho}\sum_{j,j^{\prime}}\int\frac{dk}{2\pi}\int\frac{dk^{\prime}}{2\pi}\widetilde{h}_{1,j,j^{\prime}}(k,k^{\prime})e^{iz(k-k^{\prime})}\nonumber\\
&\times e^{i\phi(j-j^{\prime})}
\tfrac{jK_{j}(|k|\rho)\partial_{\rho}K_{j^{\prime}}(|k^{\prime}|\rho)-j^{\prime}\partial_{\rho}K_{j}(|k|\rho)K_{j^{\prime}}(|k^{\prime}|\rho)}{K_{j}(|k|a)K_{j^{\prime}}(|k^{\prime}|a)},\\
\mathcal{I}_{\rho z}=& -\frac{2i}{a}\sum_{j,j^{\prime}}\int\frac{dk}{2\pi}\int\frac{dk^{\prime}}{2\pi}\widetilde{h}_{1,j,j^{\prime}}(k,k^{\prime})e^{iz(k-k^{\prime})}\nonumber\\
&\times e^{i\phi(j-j^{\prime})}
\tfrac{kK_{j}(|k|\rho)\partial_{\rho}K_{j^{\prime}}(|k^{\prime}|\rho)-k^{\prime}\partial_{\rho}K_{j}(|k|\rho)K_{j^{\prime}}(|k^{\prime}|\rho)}{K_{j}(|k|a)K_{j^{\prime}}(|k^{\prime}|a)},\\
\mathcal{I}_{\phi z}=& -\frac{2}{a\rho}\sum_{j,j^{\prime}}\int\frac{dk}{2\pi}\int\frac{dk^{\prime}}{2\pi}\widetilde{h}_{1,j,j^{\prime}}(k,k^{\prime})\left(k^{\prime}j+kj^{\prime}\right)\nonumber\\
&\times e^{iz(k-k^{\prime})} e^{i\phi(j-j^{\prime})}\frac{K_{j}(|k|\rho)}{K_{j}(|k|a)}\frac{K_{j^{\prime}}(|k^{\prime}|\rho)}{K_{j^{\prime}}(|k^{\prime}|a)}.
\label{eq:i-phi-z}
\end{align}
%
with $ \widetilde{h}_{1,j,j^{\prime}} $ given by Eq. \eqref{eq:func-h-infinit} with $m=1$.
Note that the lateral force that acts on the particle arises due to the dependence of $U^{(1)}_\text{vdW}$ on $\phi$ and $z$.
Since we are interested only in the behavior of this force, hereafter we focus our attention only on $U^{(1)}_\text{vdW}$.

\subsection{Interaction energy for a cylinder with a sinusoidal corrugation along the $z$-direction}
\label{cap-3-sec-3}

Let us consider a sinusoidal corrugation profile described by Eq. \eqref{eq:def-corrug-senoid-z}.
For this situation, $U^{(1)}_\text{vdW}$ can be written as
\begin{equation}
U_{\text{vdW, cc-\ensuremath{z}}}^{\left(1\right)}\left(\mathbf{r}_{0}\right)=\frac{\delta}{8\pi\epsilon_{0}}A_{\text{cc-\ensuremath{z}}}\cos\left(k_{c} z_0-\Delta_{\text{cc-\ensuremath{z}}}\right),
\label{eq:u1-senoidal-z}
\end{equation}
where $A_{\text{cc-\ensuremath{z}}}=\sqrt{B_{\text{cc-\ensuremath{z}}}^{2}+ C_{\text{cc-\ensuremath{z}}}^{2}}$, and $\Delta_{\text{cc-\ensuremath{z}}}$ is a nontrivial phase function defined by
\begin{equation}
\Delta_{\text{cc-\ensuremath{z}}}=\arctan\left(\frac{B_{\text{cc-\ensuremath{z}}}}{C_{\text{cc-\ensuremath{z}}}}\right),
\label{eq:delta-cc-z}
\end{equation}
with
\begin{align}
B_{\text{cc-\ensuremath{z}}} & =\left\langle \hat{d}_{\rho}\hat{d}_{z}\right\rangle {\cal R}_{\rho z}^{\text{cc-\ensuremath{z}}}, \label{eq:b-cc-z} \\
C_{\text{cc-\ensuremath{z}}} & =\left\langle \hat{d}_{\rho}^{2}\right\rangle \mathcal{R}_{\rho\rho}^{\text{cc-\ensuremath{z}}}+\left\langle \hat{d}_{\phi}^{2}\right\rangle \mathcal{R}_{\phi\phi}^{\text{cc-\ensuremath{z}}}+\left\langle \hat{d}_{z}^{2}\right\rangle \mathcal{R}_{zz}^{\text{cc-\ensuremath{z}}},
\label{eq:c-cc-z} 
\end{align}
and the functions ${\cal R}_{ij}^{\text{cc-\ensuremath{z}}}$ being defined by
\begin{align}
\mathcal{R}_{\rho\rho}^{\text{cc-\ensuremath{z}}}=&-\frac{4}{\pi a}\sum_{j=0}^{\infty}{}^{\prime}\int_{0}^{\infty}dk\frac{\partial_{\rho_0}K_{j}\left(\left|k-\frac{k_{c}}{2}\right|\rho_0\right)}{K_{j}\left(\left|k-\frac{k_{c}}{2}\right|a\right)}\nonumber\\
&\times\frac{\partial_{\rho_0}K_{j}\left(\left|k+\frac{k_{c}}{2}\right|\rho_0\right)}{K_{j}\left(\left|k+\frac{k_{c}}{2}\right|a\right)}, \\
\mathcal{R}_{\phi\phi}^{\text{cc-\ensuremath{z}}}=&-\frac{4}{\pi a}\frac{1}{\rho^{2}_0}\sum_{j=1}^{\infty}\int_{0}^{\infty}dkj^{2}\frac{K_{j}\left(\left|k-\frac{k_{c}}{2}\right|\rho_0\right)}{K_{j}\left(\left|k-\frac{k_{c}}{2}\right|a\right)}\nonumber\\
&\times\frac{K_{j}\left(\left|k+\frac{k_{c}}{2}\right|\rho_0\right)}{K_{j}\left(\left|k+\frac{k_{c}}{2}\right|a\right)}, \\
\mathcal{R}_{zz}^{\text{cc-\ensuremath{z}}}=&-\frac{4}{\pi a}\sum_{j=0}^{\infty}{}^{\prime}\int_{0}^{\infty}dk\left[k^{2}-\left(\frac{k_{c}}{2}\right)^{2}\right]\nonumber\\
&\times\frac{K_{j}\left(\left|k-\frac{k_{c}}{2}\right|\rho_0\right)}{K_{j}\left(\left|k-\frac{k_{c}}{2}\right|a\right)}\frac{K_{j}\left(\left|k+\frac{k_{c}}{2}\right|\rho_0\right)}{K_{j}\left(\left|k+\frac{k_{c}}{2}\right|a\right)}, \\
\mathcal{R}_{\rho z}^{\text{cc-\ensuremath{z}}}=&-\frac{4}{\pi a}\sum_{j=0}^{\infty}{}^{\prime}\int_{0}^{\infty}dkk\frac{K_{j}\left(\left|k\right|\rho_0\right)}{K_{j}\left(\left|k\right|a\right)}
\nonumber\\
&\times\left[\frac{\partial_{\rho_0}K_{j}\left(\left|k+k_{c}\right|\rho_0\right)}{K_{j}\left(\left|k+k_{c}\right|a\right)}-\frac{\partial_{\rho_0}K_{j}\left(\left|k-k_{c}\right|\rho_0\right)}{K_{j}\left(\left|k-k_{c}\right|a\right)}\right].
\label{eq:R-rho-z}
\end{align}

From Eq. \eqref{eq:u1-senoidal-z}, considering the behavior of $U^{(1)}$ with respect to $z_0$, one can see that the stable equilibrium points of $U^{(1)}$ can be over the corrugation peaks $(\Delta_{\text{cc-\ensuremath{z}}} = \pi)$, valleys $(\Delta_{\text{cc-\ensuremath{z}}} = 0)$ or over intermediate points between a peak and a valley $(\Delta_{\text{cc-\ensuremath{z}}} \neq 0, \pi)$.
Such possibilities were called in Ref. \cite{Nogueira-PRA-2021} as peak, valley and intermediate regimes, respectively.
In Refs. \cite{Queiroz-PRA-2021,Nogueira-PRA-2022}, we showed how modifications on the electromagnetic properties of the surface affect the occurrence of these regimes.
Otherwise, with Eqs. \eqref{eq:u1-senoidal-z}-\eqref{eq:R-rho-z}, we can now study how modifications on the curvature of the corrugated surface affect the occurrence of these regimes.
For this, let us consider a class of cylindrically symmetric polarizable particles, whose tensor $ \langle \hat{d}_i \hat{d}_j \rangle $ diagonalized, in cylindrical coordinates, is represented by $ \langle\hat{d}_{i}\hat{d}_{j}\rangle=\text{diag}\left(\langle\hat{d}_{n}^{2}\rangle,\langle\hat{d}_{n}^{2}\rangle,\langle\hat{d}_{p}^{2}\rangle\right) $, with $ \langle\hat{d}_{p}^{2}\rangle > \langle\hat{d}_{n}^{2}\rangle $.
For a general orientation of this particle, the components of $ \langle \hat{d}_i \hat{d}_j \rangle $ in terms of the spherical angles $(\theta,\varphi)$, with respect to the coordinates system $(\rho,\phi,z)$, are given by:
\begin{align}
\label{eq:drho2}
\langle\hat{d}_{\rho}^{2}\rangle & =\langle\hat{d}_{p}^{2}\rangle\left[\beta+\left(1-\beta\right)\sin^{2}\theta\cos^{2}\varphi\right],\\
\langle\hat{d}_{\phi}^{2}\rangle & =\langle\hat{d}_{p}^{2}\rangle\left[\beta+\left(1-\beta\right)\sin^{2}\theta\sin^{2}\varphi\right],\\
\langle\hat{d}_{z}^{2}\rangle & =\langle\hat{d}_{p}^{2}\rangle\left[\beta+\left(1-\beta\right)\cos^{2}\theta\right],\\
\langle\hat{d}_{\rho}\hat{d}_{\phi}\rangle & =\langle\hat{d}_{p}^{2}\rangle\frac{\left(1-\beta\right)}{2}\sin2\varphi\sin^{2}\theta,\\
\langle\hat{d}_{\rho}\hat{d}_{z}\rangle & =\langle\hat{d}_{p}^{2}\rangle\frac{\left(1-\beta\right)}{2}\sin2\theta\cos\varphi,\\
\langle\hat{d}_{\phi}\hat{d}_{z}\rangle & =\langle\hat{d}_{p}^{2}\rangle\frac{\left(1-\beta\right)}{2}\sin2\theta\sin\varphi,
\label{eq:dphi-dz}
\end{align}
with $\beta=\langle\hat{d}_{n}^{2}\rangle/\langle\hat{d}_{p}^{2}\rangle$.

\begin{figure}
\centering
\epsfig{file=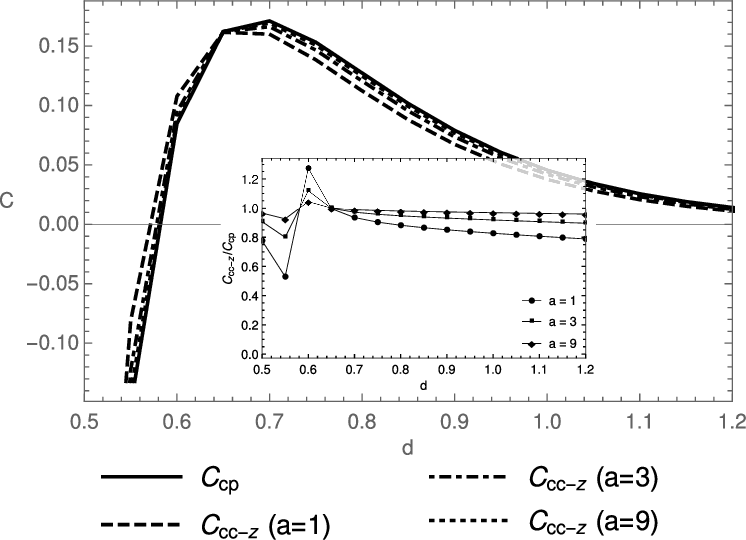, width=.7 \linewidth}
\caption{
Behavior of $C_{\text{cc-\ensuremath{z}}}$ and $C_{\text{cp}}$ (solid line) versus $ d $ for a particle oriented such that $\theta=0$ (its axis is parallel to the $z$-axis).
We consider $ \beta = 0.2 $, $ k_c = 6 $ and $ a = 1 $ (dashed line), $ a = 3 $ (dot-dashed line), and $ a = 9 $ (dotted line).
Using the same values for $ \beta $ and $ k_c $, the inset shows the behavior of the ratio $C_{\text{cc-\ensuremath{z}}} / C_{\text{cp}}$ versus $ d $ for some values of $ a $.
}
\label{fig:regime-pico-vale-z}
\end{figure}

Let us start analyzing the curvature effects on the occurrence of the peak and valley regimes.
As mentioned above, we have peak and valley regimes when  $\Delta_{\text{cc-\ensuremath{z}}} = \pi$, and $\Delta_{\text{cc-\ensuremath{z}}} = 0$, respectively. 
In this way, from Eqs. \eqref{eq:delta-cc-z}-\eqref{eq:c-cc-z}, one sees that such regimes can only occur when $B_{\text{cc-\ensuremath{z}}}=0$, which occurs when $\langle\hat{d}_{\rho}\hat{d}_{z}\rangle = 0 $ (the particle is oriented such that its axis coincides with $\rho$, $\phi$, or $z$).
In addition, we have one or the other of these regimes depending on the sign of $C_{\text{cc-\ensuremath{z}}}$, where we have the valley regime when it is positive, and the peak regime when it is negative [see Eq. \eqref{eq:delta-cc-z}].
By considering the particle oriented such that its axis is parallel to the $z$-axis $(\theta=0)$, in Fig. \ref{fig:regime-pico-vale-z} we show the behaviors of $C_{\text{cc-\ensuremath{z}}}$ and the corresponding function (named $C_{\text{cp}}$) for the case of a corrugated plane, which was calculated in Ref. \cite{Nogueira-PRA-2021}.
In Fig. \ref{fig:regime-pico-vale-z}, one notes that, for the case of a corrugation occurring in the $ z $-direction, the presence of curvature on the corrugated surface has a small effect on the occurrence of the valley regime, since $C_{\text{cc-\ensuremath{z}}}$, when compared with $C_{\text{cp}}$, changes its sign by a slightly different value of $ d $ for different values of $ a $. 
One can also see that, as $ a $ increases, the behavior of $C_{\text{cc-\ensuremath{z}}}$, for a given value of $ d $, approaches to the behavior of $C_{\text{cp}}$, which means that the curvature effects vanish, as expected.
This is reinforced in the inset, where we show the behavior of the ratio $C_{\text{cc-\ensuremath{z}}}/C_{\text{cp}}$ in terms of $ d $ (the sudden oscillation of the curves between $ d=0.54 $ and $ d=0.6 $ occurs due to the change of sign of the functions $C_{\text{cc-\ensuremath{z}}}$ and $C_{\text{cp}}$, which occurs in this interval).

\begin{figure}
\centering
\epsfig{file=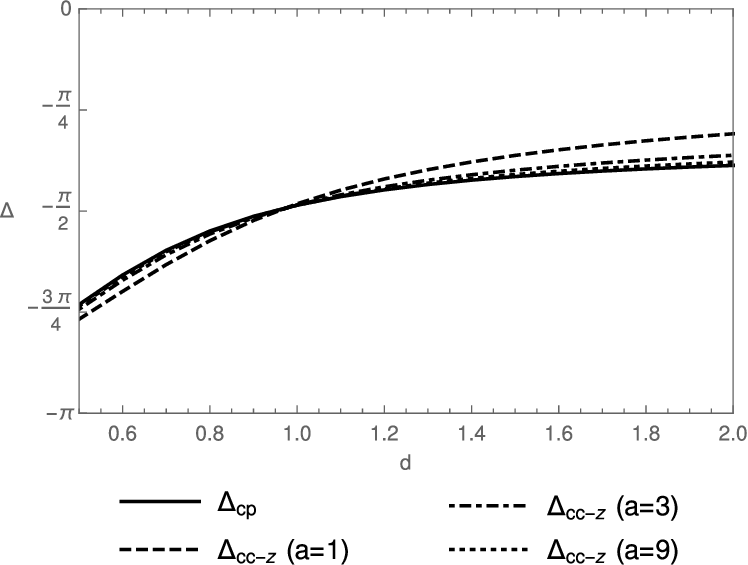, width=0.7 \linewidth}
\caption{
Behavior of $\Delta_{\text{cc-\ensuremath{z}}}$ and $\Delta_{\text{cp}}$ (solid line) versus $ d $ for a particle oriented with $\theta=\pi/6$  and $\varphi=0$.
We consider $ \beta = 0.2 $, $ k_c = 6 $ and $ a = 1 $ (dashed line), $ a = 3 $ (dot-dashed line), and $ a = 9 $ (dotted line).
}
\label{fig:regime-intermediario-z}
\end{figure}

Now, let us analyze the curvature effects on the occurrence of the intermediate regimes.
As mentioned above, we have intermediate regimes when  $\Delta_{\text{cc-\ensuremath{z}}} \neq 0, \pi$, which can only occur when $B_{\text{cc-\ensuremath{z}}} \neq 0$, which occurs when $\langle\hat{d}_{\rho}\hat{d}_{z}\rangle \neq 0 $ (the particle is oriented such that its axis do not coincide with $\rho$, $\phi$, or $z$).
By considering the particle oriented with $\theta=\pi/6$  and $\varphi=0$, in Fig. \ref{fig:regime-intermediario-z} we show the behaviors of $\Delta_{\text{cc-\ensuremath{z}}}$ and $\Delta_{\text{cp}}$.
In Fig. \ref{fig:regime-intermediario-z}, one notes that, for the case of a corrugation occurring in the $ z $-direction, the presence of curvature on the corrugated surface can inhibit the occurrence of the intermediate regimes up to a certain value of $ d $, above which it begins to amplify the occurrence of the effect.
Lastly, as expected, $\Delta_{\text{cc-\ensuremath{z}}}$ approaches to $\Delta_{\text{cp}}$ as $ a $ increases, meaning that the curvature effect decreases by increasing $ a $.

\subsection{Interaction energy for a cylinder with a sinusoidal corrugation along the $\phi$-direction}
\label{cap-3-sec-3-subsec-2}

Let us now consider a sinusoidal corrugation profile described by Eq. \eqref{eq:def-corrug-senoid}.
For this situation, $U^{(1)}_\text{vdW}$ can be written as
\begin{equation}
U_{\text{vdW, cc-\ensuremath{\phi}}}^{\left(1\right)}\left(\mathbf{r}_{0}\right)=\frac{\delta}{8\pi\epsilon_{0}}A_{\text{cc-\ensuremath{\phi}}}\cos\left(k_{c}a\phi_{0}-\Delta_{\text{cc-\ensuremath{\ensuremath{\phi}}}}\right),
\label{eq:u1-senoidal-phi}
\end{equation}
where $A_{\text{cc-\ensuremath{\phi}}}=\sqrt{B_{\text{cc-\ensuremath{\phi}}}^{2}+C_{\text{cc-\ensuremath{\phi}}}^{2}}$, and $\Delta_{\text{cc-\ensuremath{\phi}}}$ is a nontrivial phase function defined by
\begin{equation}
\Delta_{\text{cc-\ensuremath{\phi}}}=\arctan\left(\frac{B_{\text{cc-\ensuremath{\phi}}}}{C_{\text{cc-\ensuremath{\phi}}}}\right),
\label{eq:delta-cc-phi}
\end{equation}
with
\begin{align}
\label{eq:b-cc-phi}
B_{\text{cc-\ensuremath{\phi}}} & =\left\langle \hat{d}_{\rho}\hat{d}_{\phi}\right\rangle \mathcal{R}_{\rho\phi}^{\text{cc-\ensuremath{\phi}}},\\
C_{\text{cc-\ensuremath{\phi}}} & =\left\langle \hat{d}_{\rho}^{2}\right\rangle \mathcal{R}_{\rho\rho}^{\text{cc-\ensuremath{\phi}}}+\left\langle \hat{d}_{\phi}^{2}\right\rangle \mathcal{R}_{\phi\phi}^{\text{cc-\ensuremath{\phi}}}+\left\langle \hat{d}_{z}^{2}\right\rangle \mathcal{R}_{zz}^{\text{cc-\ensuremath{\phi}}},
\label{eq:c-cc-phi}
\end{align}
and the functions ${\cal R}_{ij}^{\text{cc-\ensuremath{\phi}}}$ being defined by
\begin{align}
\mathcal{R}_{\rho\rho}^{\text{cc-\ensuremath{\phi}}} =&-\frac{2}{\pi a}\int_{0}^{\infty}dk\sum_{j=0}^{\infty}{}^{\prime}\frac{\partial_{\rho_{0}}K_{j}\left(\left|k\right|\rho_{0}\right)}{K_{j}\left(\left|k\right|a\right)}\nonumber\\
&\times\left[\frac{\partial_{\rho_{0}}K_{j+k_{c}a}\left(\left|k\right|\rho_{0}\right)}{K_{j+k_{c}a}\left(\left|k\right|a\right)}+\frac{\partial_{\rho_{0}}K_{j-k_{c}a}\left(\left|k\right|\rho_{0}\right)}{K_{j-k_{c}a}\left(\left|k\right|a\right)}\right],
\end{align}
\begin{gather}
%
\mathcal{R}_{\phi\phi}^{\text{cc-\ensuremath{\phi}}}=-\frac{2}{\pi a\rho_{0}^{2}}\int_{0}^{\infty}dk\sum_{j=1}^{\infty}j\frac{K_{j}(|k|\rho_{0})}{K_{j}(|k|a)}\nonumber\\
\left[(j+k_{c}a)\frac{K_{j+k_{c}a}(|k|\rho_{0})}{K_{j+k_{c}a}(|k|a)}\right.
+\left.(j-k_{c}a)\frac{K_{j-k_{c}a}(|k|\rho_{0})}{K_{j-k_{c}a}(|k|a)}\right],
\end{gather}
\begin{align}
\mathcal{R}_{zz}^{\text{cc-\ensuremath{\phi}}}=&-\frac{2}{\pi a}\int_{0}^{\infty}dk\sum_{j=0}^{\infty}{}^{\prime}k^{2}\frac{K_{j}(|k|\rho_{0})}{K_{j}(|k|a)}\nonumber\\
&\times\left[\frac{K_{j+k_{c}a}(|k|\rho_{0})}{K_{j+k_{c}a}(|k|a)}+\frac{K_{j-k_{c}a}(|k|\rho_{0})}{K_{j-k_{c}a}(|k|a)}\right],
\end{align}
\begin{align}
\mathcal{R}_{\rho\phi}^{\text{cc-\ensuremath{\phi}}}=&-\frac{4}{\pi a\rho_{0}}\int_{0}^{\infty}dk\sum_{j=1}^{\infty}\frac{jK_{j}\left(\left|k\right|\rho_{0}\right)}{K_{j}\left(\left|k\right|a\right)}
\nonumber\\
&\times\left[\frac{\partial_{\rho_{0}}K_{j+k_{c}a}\left(\left|k\right|\rho_{0}\right)}{K_{j+k_{c}a}\left(\left|k\right|a\right)}-\frac{\partial_{\rho_{0}}K_{j-k_{c}a}\left(\left|k\right|\rho_{0}\right)}{K_{j-k_{c}a}\left(\left|k\right|a\right)}\right].
\label{eq:R-rho-phi}
\end{align}
Similar to the case of a corrugation occurring in the $z$-direction, from Eq. \eqref{eq:u1-senoidal-phi}, considering the behavior of $U^{(1)}$ with respect to $\phi_0$, one can see that the stable equilibrium points of $U^{(1)}$ can be over the corrugation peaks $(\Delta_{\text{cc-\ensuremath{\phi}}} = \pi)$, valleys $(\Delta_{\text{cc-\ensuremath{\phi}}} = 0)$ or over intermediate points between a peak and a valley $(\Delta_{\text{cc-\ensuremath{\phi}}} \neq 0, \pi)$.
In other words, we can have peak, valley and intermediate regimes, depending on the value of $ \Delta_{\text{cc-\ensuremath{\phi}}} $.
With Eqs. \eqref{eq:u1-senoidal-phi}-\eqref{eq:R-rho-phi}, we can also study how modifications on the curvature of the corrugated surface affect the occurrence of these regimes.
For this, let us consider again cylindrically symmetric polarizable particles, whose components of the tensor $ \langle \hat{d}_i \hat{d}_j \rangle $ for a general orientation of this particle are given in Eqs. \eqref{eq:drho2}-\eqref{eq:dphi-dz}.

\begin{figure}
\centering
\epsfig{file=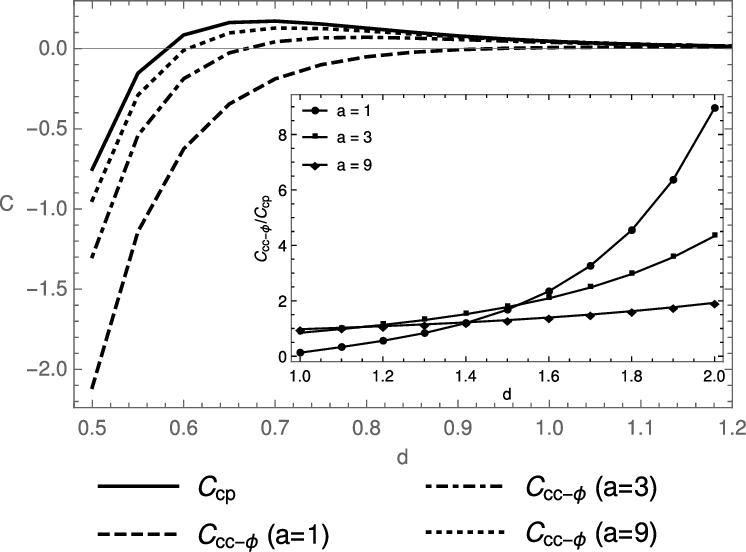, width=.7 \linewidth}
\caption{
Behavior of $C_{\text{cc-\ensuremath{\phi}}}$ and $C_{\text{cp}}$ (solid line) versus $ d $ for a particle oriented such that $\theta=\pi/2$ and $ \varphi=\pi/2 $ (its axis is parallel to the $\phi$-axis).
We consider $ \beta = 0.2 $, $ k_c = 6 $ and $ a = 1 $ (dashed line), $ a = 3 $ (dot-dashed line), and $ a = 9 $ (dotted line).
Using the same values for $ \beta $ and $ k_c $, the inset shows the behavior of the ratio $C_{\text{cc-\ensuremath{\phi}}} / C_{\text{cp}}$ versus $ d $ for some values of $ a $.
}
\label{fig:regime-pico-vale-phi}
\end{figure}

Let us start analyzing the curvature effects on the occurrence of the peak and valley regimes.
As mentioned above, we have peak and valley regimes when  $\Delta_{\text{cc-\ensuremath{\phi}}} = \pi$, and $\Delta_{\text{cc-\ensuremath{\phi}}} = 0$, respectively. 
In this way, from Eqs. \eqref{eq:delta-cc-phi}-\eqref{eq:c-cc-phi}, one sees that such regimes can only occur when $B_{\text{cc-\ensuremath{\phi}}}=0$, which occurs when $\langle\hat{d}_{\rho}\hat{d}_{\phi}\rangle = 0 $ (the particle is oriented such that its axis coincides with $\rho$, $\phi$, or $z$).
In addition, we have one or the other of these regimes depending on the sign of $C_{\text{cc-\ensuremath{\phi}}}$, where we have the valley regime when it is positive, and the peak regime when it is negative [see Eq. \eqref{eq:delta-cc-phi}].
By considering the particle oriented such that its axis is parallel to the $\phi$-axis $(\theta=\pi/2,\varphi=\pi/2)$, in Fig. \ref{fig:regime-pico-vale-phi} we show the behaviors of $C_{\text{cc-\ensuremath{\phi}}}$ and $C_{\text{cp}}$.
In Fig. \ref{fig:regime-pico-vale-phi}, one notes that, for the case of a corrugation occurring in the $ \phi $-direction, the presence of curvature on the corrugated surface inhibits the occurrence of the valley regime, since $C_{\text{cc-\ensuremath{\phi}}}$, when compared with $C_{\text{cp}}$, changes its sign at higher values of $ d $ by decreasing $ a $. 
Such behavior can be related to the fact that, in this case, the curvature of the cylinder occurs in the same direction as the oscillations of the corrugation, i.e, modifications on the curvature of the cylinder modify the structure of the corrugation itself.
Although this, one can also see that, as $ a $ increases, the behavior of $C_{\text{cc-\ensuremath{\phi}}}$, for a given value of $ d $, approaches to the behavior of $C_{\text{cp}}$, which means that the curvature effects vanish, as expected.
On the other hand, as $ d $ increases, the curvature effects become higher, which can be better seen from the inset of Fig. \ref{fig:regime-pico-vale-phi}, where we show the behavior of the ratio $C_{\text{cc-\ensuremath{\phi}}}/C_{\text{cp}}$ in terms of $ d $.

\begin{figure}
\centering
\epsfig{file=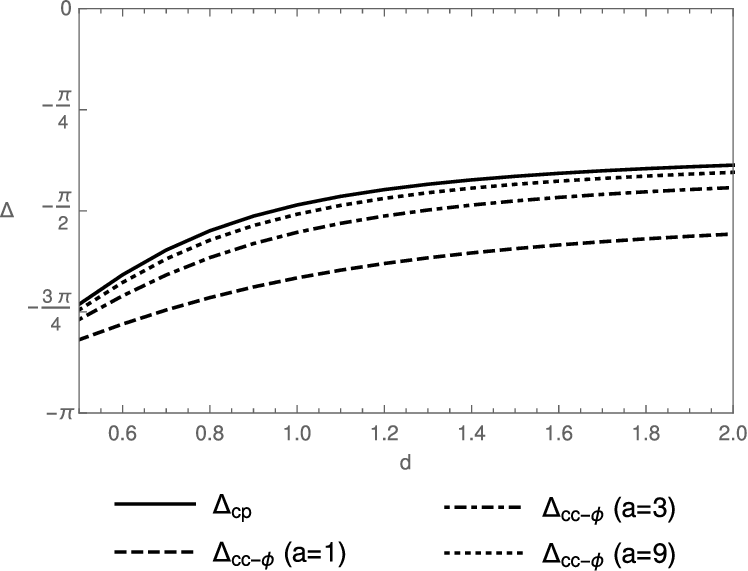, width=0.7 \linewidth}
\caption{
Behavior of $\Delta_{\text{cc-\ensuremath{\phi}}}$ and $\Delta_{\text{cp}}$ (solid line) versus $ d $ for a particle oriented with $\theta=\pi/2$  and $\varphi=\pi/3$.
We consider $ \beta = 0.2 $, $ k_c = 6 $ and $ a = 1 $ (dashed line), $ a = 3 $ (dot-dashed line), and $ a = 9 $ (dotted line).
}
\label{fig:regime-intermediario-phi}
\end{figure}

Now, let us analyze the curvature effects on the occurrence of the intermediate regimes.
As mentioned above, we have intermediate regimes when  $\Delta_{\text{cc-\ensuremath{\phi}}} \neq 0, \pi$, which can only occur when $B_{\text{cc-\ensuremath{\phi}}} \neq 0$, which occurs when $\langle\hat{d}_{\rho}\hat{d}_{\phi}\rangle \neq 0 $ (the particle is oriented such that its axis do not coincide with $\rho$, $\phi$, or $z$).
By considering the particle oriented with $\theta=\pi/2$  and $\varphi=\pi/3$, in Fig. \ref{fig:regime-intermediario-phi} we show the behaviors of $\Delta_{\text{cc-\ensuremath{\phi}}}$ and $\Delta_{\text{cp}}$.
In Fig. \ref{fig:regime-intermediario-phi}, one notes that, for the case of a corrugation occurring in the $ \phi $-direction, the presence of curvature on the corrugated surface also inhibit the occurrence of the intermediate regimes by decreasing $ a $.
Otherwise, as expected, $\Delta_{\text{cc-\ensuremath{\phi}}}$ approaches to $\Delta_{\text{cp}}$ as $ a $ increases, meaning that the curvature effect vanishes for $ a \gg d $.
%
%
%
%

%
\section{Final Remarks}
\label{sec-final-regimes-curvatura}

We studied how the peak, valley and intermediate regimes for the lateral vdW force are affected by modifications on the geometric properties of the surface itself on which the corrugation occurs.
Our main results are given by Eqs. \eqref{eq:u-corrug-general}-\eqref{eq:i-phi-z}, which, applied to a sinusoidal corrugation [Eqs. \eqref{eq:u1-senoidal-z} or \eqref{eq:u1-senoidal-phi}], showed that, in general, the presence of curvature on the corrugated surface inhibits the occurrence of valley and intermediate regimes.
When the corrugation occurs in the $ z $-direction [Eq. \eqref{eq:u1-senoidal-z}], we showed that the presence of curvature on the corrugated surface has a small effect on the occurrence of the valley regime, practically not affecting it.
On the other hand, for the intermediate regimes, the presence of curvature inhibit their occurrence up to a certain particle-surface distance, above which it begins to amplify the occurrence of the effect.
When the corrugation occurs in the $ \phi $-direction [Eq. \eqref{eq:u1-senoidal-phi}], we showed that, different from the previous case, the presence of curvature on the corrugated surface inhibits the occurrence of the valley regime.
We argued that such behavior can be related to the fact that, in this case, by modifying the curvature of the cylinder we modify the structure of the corrugation itself.
For the intermediate regimes, we showed that the presence of curvature on the corrugated surface also inhibit their occurrence.

\begin{acknowledgments}
	The authors thank C. Farina and V. S. Alves for valuable discussions.
	A.P.C and L.Q. were supported by the Coordena\c{c}\~{a}o de Aperfei\c{c}oamento de Pessoal de N\'{i}vel Superior - Brasil (CAPES), Finance Code 001.
	This work was partially supported by CNPq - Brazil, Processo 408735/2023-6 CNPq/MCTI.
	\end{acknowledgments}
%


%

\end{document}